\documentclass[twocolumn,showpacs,preprintnumbers,amsmath,amssymb]{revtex4-2}

\usepackage[utf8]{inputenc}
\usepackage[pdftex]{graphicx}
\usepackage{caption}
\usepackage{subcaption}
\usepackage{mathrsfs}
\usepackage[colorlinks, breaklinks, urlcolor={blue}, linkcolor={red}, citecolor={blue}]{hyperref}
\usepackage{array}
\usepackage{amsmath}
\usepackage{type1cm}
\usepackage{lettrine}
\usepackage[english]{babel}
\usepackage{lmodern}
\usepackage{microtype}
\usepackage{booktabs}
\usepackage[T1]{fontenc}
\usepackage[boxed, vlined]{algorithm2e}
\usepackage{caption}
\usepackage{braket}
\usepackage{xcolor}
\usepackage{orcidlink}
\usepackage{bm}
\usepackage{bbold}
\usepackage{upgreek}

\begin{document}

\title{Bell's theorem in time without inequalities}

\author{Md.~Manirul Ali \orcidlink{0000-0002-5076-7619}}
\email{manirul@citchennai.net}
\affiliation{Centre for Quantum Science and Technology, Chennai Institute of Technology, Chennai 600069, India}

\begin{abstract}
Bell's theorem revealed that a local hidden-variable model cannot completely reproduce the quantum mechanical predictions. Bell's inequality provides an upper bound under the locality and reality assumptions that can be violated by correlated measurement statistics of quantum mechanics. Greenberger, Horne, and Zeilinger (GHZ) gave a more compelling proof of Bell's theorem without inequalities by considering perfect correlations rather than statistical correlations. This work presents a temporal analog of the GHZ argument that establishes {\it Bell's theorem in time without inequalities}.
\end{abstract}


\maketitle

Einstein, Podolski, and Rosen (EPR) in 1935 remarked \cite{EPR} that the description of nature depicted by quantum mechanics is
incomplete, whereas a more complete realistic theory is possible where physical observables can have predetermined definite
values even in the absence of measurement. In realistic hidden-variables models \cite{Bell66}, the very occurrence of a definite
outcome in an individual measurement of an observable is predetermined by some hidden-variables. Bell's theorem showed
\cite{Bell64} that a local hidden-variables model cannot completely reproduce some spatially-separated measurement statistics
or statistical correlations of a pair of two-level quantum systems. Motivated by Bell's theorem, a temporal analogue of Bell
inequalities was provided by Leggett and Garg \cite{Leggett85} in terms of time-separated measurement statistics or temporal
statistical correlation functions. Leggett-Garg inequality was constructed under two classical assumptions (a1) macrorealism:
macroscopic systems are always in a definite state with well-defined pre-existing value and (a2) noninvasive measurability: that
pre-existing value can be measured without disturbing the system. These classical realism and noninvasive measurability assumptions
together allow us to consider a single joint probability distribution for all time-separated measurement observables in an
experiment, from which the marginal probability distributions can also be obtained \cite{Leggett85,Unified13,Unified14,RevNori14}.
Hence, Bell-inequalities and Leggett-Garg inequalities have a common origin, they are related by their implicit assumption of joint
realism that there exists a classical joint probability distribution associated to the joint measurements that determines all measurement
statistics \cite{Leggett85,Unified13,Unified14,RevNori14}. More precisely, the common assumption is that there exists a classical
probability distribution that can reproduce all spatially-separated or time-separated measurement statistics in quantum mechanics.
Bell-inequalities and the Leggett-Garg inequalities both were shown to be violated by certain quantum mechanical statistical
predictions \cite{BellExpt1,BellExpt2,BellExpt3,BellExpt4,BellExpt5,LGExpt1,LGExpt2,LGExpt3,LGExpt4,LGExpt5,LGExpt6}.

A more powerful demonstration of Bell's theorem was given by Greenberger, Horne, and Zeilinger
\cite{GHZ89,GHZ90,Mermin90} using three or more correlated spin-1/2 quantum particles.
Unlike Bell's original theorem, GHZ analysis examines perfect correlations rather than
statistical correlations or imperfect correlations, and it is independent of any inequality bound. Hence GHZ-nonlocality tests Bell's local
realism hypothesis without inequalities. To test Bell inequalities or Leggett-Garg inequalities one needs to repeat the measurements
many times with the access of many copies of the state, to estimate the spatial or temporal correlation functions using the probabilities
of measurement outcomes, whereas the GHZ experiment needs only be repeated once in principle. In this work, we are providing a
temporal analogue of GHZ argument which is surprisingly unexplored in the literature. GHZ nonlocality argument tests Bell's local
realism without inequalities, in the same spirit our result in this work tests Leggett-Garg's classical realism and noninvasive measurability
hypothesis without any statistical correlation functions or inequalities. We provide here a non-probabilistic argument. Leggett-Garg
inequalities and Bell-inequalities have identical structures and are intimately related. Bell-inequalities provide upper bounds on
correlations between measurements on spatially separated systems, whereas Leggett-Garg inequalities provide upper bounds on
correlations of measurements at different times. If one perceives Leggett-Garg inequality as the temporal version of Bell-inequality
\cite{LGExpt1,Paz1993,Ruskov06,Emary12,Souza11}, then our result or proof can be regarded as a stronger test of
{\it Bell’s theorem in time without a statistical inequality}.

The GHZ experiment \cite{GHZexp1,GHZexp2,GHZexp3} works as follows. Consider three spin-1/2 particles prepared in an
entangled state
\begin{eqnarray}
|\Psi_{GHZ}\rangle = \frac{1}{\sqrt{2}} \left(|\uparrow\rangle_1 |\uparrow\rangle_2 |\uparrow\rangle_3
+ |\downarrow\rangle_1 |\downarrow\rangle_2 |\downarrow\rangle_3 \right),
\label{ghz}
\end{eqnarray}
where $|\uparrow\rangle$ and $|\downarrow\rangle$ represent the eigenstates of spin measurement operator $\sigma_z$
along z-axis with eigenvalues $+1$ and $-1$ respectively. Let us now consider four mutually commuting Hermitian
operators in the three-qubit Hilbert space as
\begin{eqnarray}
\label{gxxx}
& \sigma_x^1~\sigma_x^2~\sigma_x^3, \\
\label{gxyy}
& \sigma_x^1~\sigma_y^2~\sigma_y^3, \\
\label{gyyx}
& \sigma_y^1~\sigma_y^2~\sigma_x^3, \\
\label{gyxy}
& \sigma_y^1~\sigma_x^2~\sigma_y^3.
\end{eqnarray}
These four operators commute with each other for which they can have simultaneous eigenstates and eigenvalues.
The state $|\Psi_{GHZ}\rangle$ is an eigenstate of the operators $\sigma_x^1 \sigma_y^2 \sigma_y^3$,
$\sigma_y^1 \sigma_y^2 \sigma_x^3$, and $\sigma_y^1 \sigma_x^2 \sigma_y^3$ with a common eigenvalue $-1$.
One can also verify that $|\Psi_{GHZ}\rangle$ is an eigenstate of $\sigma_x^1 \sigma_x^2 \sigma_x^3$
with eigenvalue $+1$. According to quantum mechanics, the eigenvalue equations for the four mutually commuting
Hermitian operators (\ref{gxxx}-\ref{gyxy}) are given by
\begin{eqnarray}
\label{xxxg}
\sigma_x^1~\sigma_x^2~\sigma_x^3~|\Psi_{GHZ}\rangle &=& (+1)~|\Psi_{GHZ}\rangle, \\
\label{xyyg}
\sigma_x^1~\sigma_y^2~\sigma_y^3~|\Psi_{GHZ}\rangle &=& (-1)~|\Psi_{GHZ}\rangle, \\
\label{yyxg}
 \sigma_y^1~\sigma_y^2~\sigma_x^3~|\Psi_{GHZ}\rangle &=& (-1)~|\Psi_{GHZ}\rangle, \\
\label{yxyg}
 \sigma_y^1~\sigma_x^2~\sigma_y^3~|\Psi_{GHZ}\rangle &=& (-1)~|\Psi_{GHZ}\rangle.
\end{eqnarray}
GHZ argument was that these quantum results are incompatible with local hidden-variables model.
Suppose the outcomes given by the eigenvalues of the local spin measurements $\sigma_x^1$, $\sigma_x^2$, $\sigma_x^3$,
$\sigma_y^1$, $\sigma_y^2$, $\sigma_y^3$ are fixed by some local hidden-variables model as
$v_r^i=\pm 1$, $r \in {x,y,z}$ and $i \in {1,2,3}$. Those predetermined fixed values must satisfy the relations
\begin{eqnarray}
\label{vxxx}
& v_x^1~v_x^2~v_x^3 = +1, \\
\label{vxyy}
& v_x^1~v_y^2~v_y^3 = -1, \\
\label{vyyx}
& v_y^1~v_y^2~v_x^3 = -1, \\
\label{vyxy}
& v_y^1~v_x^2~v_y^3 = -1,
\end{eqnarray}
to reproduce the quantum mechanical results given by the equations (\ref{xxxg}-\ref{yxyg}).

Now, the product of measurement outcomes for the four observables $\sigma_x^1 \sigma_x^2 \sigma_x^3$,
$\sigma_x^1 \sigma_y^2 \sigma_y^3$, $\sigma_y^1 \sigma_y^2 \sigma_x^3$, and $\sigma_y^1 \sigma_x^2 \sigma_y^3$
should be $-1$, according to quantum mechanical results (\ref{xxxg}-\ref{yxyg}).
Whereas according to the left hand sides of Eqs.~(\ref{vxxx}-\ref{vyxy}), the product of the predetermined measurement
values comes out to be $(v_x^1)^2(v_y^1)^2(v_x^2)^2(v_y^2)^2(v_x^3)^2(v_y^3)^2 = +1$. This happens because
the values $v_r^i$ are fixed under a local hidden-variables scheme and each local spin measurement along a particular
axis appears twice in the overall set of four product measurements. This is in direct contradiction with the results of quantum
mechanics. GHZ experiment demonstrates the nonlocal nature of quantum mechanics that violates our classical intuition
of local realism. Hence the results of quantum measurements on three spin-1/2 particles prepared in an entangled state
$|\Psi_{GHZ}\rangle$ cannot be explained if each of the particles are assigned predetermined values for their local spin
components before the measurement. GHZ argument provides the sharpest possible contradiction between quantum
mechanics and EPR reality criterion \cite{Mermin90}.

Next, we present a temporal version of GHZ situation where the measurements are now time-separated. Let us consider
a single spin-1/2 system evolving under a suitable Hamiltonian $H$. Analogous to the three-particle version of GHZ
argument, here we consider single-particle spin measurements at three successive times $t_1$, $t_2$, and $t_3$ with
$t_3>t_2>t_1$. We consider the following four joint measurement operators in time as
\begin{eqnarray}
\label{xxx}
& \sigma_x(t_3)~\sigma_x(t_2)~\sigma_x(t_1), \\
\label{xyy}
& \sigma_x(t_3)~\sigma_y(t_2)~\sigma_y(t_1), \\
\label{yyx}
& \sigma_y(t_3)~\sigma_y(t_2)~\sigma_x(t_1), \\
\label{yxy}
& \sigma_y(t_3)~\sigma_x(t_2)~\sigma_y(t_1).
\end{eqnarray}
Quantum mechanics prescribes that the measurement results of an observable are given by the eigenvalues of the
Hermitian operator associated to that observable. Consider an attempt to assign values (pre-existing values that
can be measured without disturbing the system) to the six individual Hermitian operators $\sigma_x(t_1)$,
$\sigma_x(t_2)$, $\sigma_x(t_3)$, $\sigma_y(t_1)$, $\sigma_y(t_2)$, $\sigma_y(t_3)$
as the corresponding eigenvalues of the operators $m_x^1$, $m_x^2$, $m_x^3$, $m_y^1$, $m_y^2$, $m_y^3$ respectively.
This attempt of value-assignment will fail as is shown below.

We consider a quantum spin-1/2 system evolving under the Hamiltonian $H = \frac{\hbar}{2} \omega \sigma_z$.
The time evolution dynamics of the Pauli spin operators are given by the Heisenberg equations of motion
\begin{eqnarray}
\label{sxh}
& \frac{d}{dt} \sigma_x(t) = \frac{1}{i\hbar} \left[ \sigma_x(t),  H \right], \\
\label{syh}
& \frac{d}{dt} \sigma_y(t) = \frac{1}{i\hbar} \left[ \sigma_y(t),  H \right],
\end{eqnarray}
where the Heisenberg picture operators are
\begin{eqnarray}
\label{sxt}
& \sigma_x(t) = e^{i H t / \hbar} ~ \sigma_x ~ e^{ - i H t / \hbar}, \\
\label{syt}
& \sigma_y(t) = e^{i H t / \hbar} ~ \sigma_y ~ e^{ - i H t / \hbar}.
\end{eqnarray}
The solutions of the Heisenberg equations of motion are given by
\begin{eqnarray}
\label{sxt}
& \sigma_x(t) = \sigma_x \cos(\omega t) - \sigma_y \sin(\omega t), \\
\label{syt}
& \sigma_y(t) = \sigma_y \cos(\omega t) + \sigma_x \sin(\omega t),
\end{eqnarray}
where
\begin{eqnarray}
\sigma_{x} = \left(\begin{array}{cc}
 0  & 1\\
1 & 0\\
\end{array}
\right),
\sigma_{y} = \left(\begin{array}{cc}
 0  & -i\\
i & 0\\
\end{array}
\right),
\sigma_{z} = \left(\begin{array}{cc}
 1  & 0\\
0 & -1\\
\end{array}
\right).
\end{eqnarray}

We consider the time-dependent dynamical observables $\sigma_x(t)$, $\sigma_y(t)$ at
three different times $t_1=0$, $t_2=\frac{\pi}{2\omega}$, and $t_3=\frac{3\pi}{2\omega}$.
Then according to the time evolution equations (\ref{sxt}) and (\ref{syt}), the four operators
given by Eqs.~(\ref{xxx}-\ref{yxy}) become Hermitian operators for the above
choice of times $t_1$, $t_2$, and $t_3$. Moreover, under that situation all the four operators (\ref{xxx}-\ref{yxy})
also commute with each other. Because they all commute, the four operators can have simultaneous eigenstates and
eigenvalues. In this situation, one can verify that the state $|\Phi\rangle=(1/\sqrt{2})(| \uparrow \rangle + | \downarrow \rangle)$
is an eigenstate of the three operators $\sigma_x(t_3)\sigma_x(t_2)\sigma_x(t_1)$, $\sigma_x(t_3)\sigma_y(t_2)\sigma_y(t_1)$,
and $\sigma_y(t_3)\sigma_y(t_2)\sigma_x(t_1)$ with a common eigenvalue $-1$. Similarly, the state $|\Phi\rangle$
is also an eigenstate of the fourth operator $\sigma_y(t_3)\sigma_x(t_2)\sigma_y(t_1)$ with eigenvalue $+1$.
Then we can write down the eigenvalue equations for the four mutually commuting Hermitian operators given by
Eqs.~(\ref{xxx}-\ref{yxy}) as follows
\begin{eqnarray}
\label{xxxe}
& \sigma_x(t_3)~\sigma_x(t_2)~\sigma_x(t_1)~|\Phi\rangle = (-1)~|\Phi\rangle, \\
\label{xyye}
& \sigma_x(t_3)~\sigma_y(t_2)~\sigma_y(t_1)~|\Phi\rangle = (-1)~|\Phi\rangle, \\
\label{yyxe}
& \sigma_y(t_3)~\sigma_y(t_2)~\sigma_x(t_1)~|\Phi\rangle = (-1)~|\Phi\rangle, \\
\label{yxye}
& \sigma_y(t_3)~\sigma_x(t_2)~\sigma_y(t_1)~|\Phi\rangle = (+1)~|\Phi\rangle.
\end{eqnarray}

We now impose the idea of classical realism$-$the results of measuring observables must have already been specified prior
to measurement. We assume that the measurement values of the individual measurement observables $\sigma_x(t_1)$, $\sigma_y(t_1)$,
$\sigma_x(t_2)$, $\sigma_y(t_2)$, $\sigma_x(t_3)$, and $\sigma_y(t_3)$ are predetermined, and measurements merely reveal those
predetermined values. Whether we perform any measurement or not, the x-component of spin $\sigma_x$ has predetermined
values $m_x^1$, $m_x^2$, and $m_x^3$ at times $t_1$, $t_2$, and $t_3$ respectively. Similarly, we assign the predetermined
measurement values of the operators $\sigma_y(t_1)$, $\sigma_y(t_2)$, $\sigma_y(t_3)$ as $m_y^1$, $m_y^2$, and $m_y^3$
respectively. Any such predetermined value is assumed to be independent of the sequence at which the sets of time-separated spin
measurements are done on this single-qubit system. This means measurement at one time does not influence the subsequent dynamics
of the system and the measurement outcomes at a later time. This is Leggett-Garg's noninvasive measurability assumption which states
that the predetermined values of the observables can be measured without disturbing the system. Now for our suitably chosen three
different times $t_1$, $t_2$ and $t_3$ and for a given initial state $|\Phi\rangle$, the four mutually commuting Hermitian operators
$\sigma_x(t_3)\sigma_x(t_2)\sigma_x(t_1)$, $\sigma_x(t_3)\sigma_y(t_2)\sigma_y(t_1)$,
$\sigma_y(t_3)\sigma_y(t_2)\sigma_x(t_1)$, and $\sigma_y(t_3)\sigma_x(t_2)\sigma_y(t_1)$ have a set of simultaneous
eigenvalues $-1$, $-1$, $-1$ and $+1$ respectively. Hence according to the requirement of quantum mechanics, consistent
with the eigenvalue equations Eqs.~(\ref{xxxe}-\ref{yxye}) we then have the following relations to satisfy
\begin{eqnarray}
\label{xxxv}
& m_x^3~m_x^2~m_x^1 = -1, \\
\label{xyyv}
& m_x^3~m_y^2~m_y^1 = -1, \\
\label{yyxv}
& m_y^3~m_y^2~m_x^1 = -1, \\
\label{yxyv}
& m_y^3~m_x^2~m_y^1 = +1.
\end{eqnarray}

In the above relations (\ref{xxxv}-\ref{yxyv}) the value of any particular quantity $m_x^1$, $m_x^2$, $m_x^3$, $m_y^1$,
$m_y^2$, or $m_y^3$ is the same irrespective of the equation in which it appears. For instance, the value of $m_x^2$ is the
same in Eqs.~(\ref{xxxv}) and (\ref{yxyv}), which is fixed according to the assumption of classical realism. However multiplying
the left hand sides of equations (\ref{xxxv}-\ref{yxyv}) yields $(m_x^1)^2(m_x^2)^2(m_x^3)^2(m_y^1)^2(m_y^2)^2(m_y^3)^2
= +1$, which is in contradiction with the product of right hand sides. The contradiction is sharp and extreme in the sense that
for some particular measurement, $+1$ result is expected whereas quantum mechanics results $-1$ instead. This result establishes
a GHZ version of {\it Bell's theorem in time without inequalities}. The {\it temporal} GHZ version of the Bell's theorem, discovered
in this work can be experimentally verified with the present quantum technology \cite{GHZexp3,SCQ,Blatt08,Kirchmair09,Zeilinger11}. Experimental realization of Greenberger-Horne-Zeilinger correlations is demonstrated in Ref.~\cite{Lloyd2000} using nuclear magnetic resonance (NMR), where the perfect correlations given by
\begin{eqnarray}
\label{xxxc}
\langle \Psi_{GHZ} \vert \sigma_x^1~\sigma_x^2~\sigma_x^3~|\Psi_{GHZ}\rangle &=& +1, \\
\label{xyyc}
\langle \Psi_{GHZ} \vert \sigma_x^1~\sigma_y^2~\sigma_y^3~|\Psi_{GHZ}\rangle &=& -1, \\
\label{yyxc}
\langle \Psi_{GHZ} \vert \sigma_y^1~\sigma_y^2~\sigma_x^3~|\Psi_{GHZ}\rangle &=& -1, \\
\label{yxyc}
\langle \Psi_{GHZ} \vert \sigma_y^1~\sigma_x^2~\sigma_y^3~|\Psi_{GHZ}\rangle &=& -1
\end{eqnarray}
were experimentally verified with respect to the GHZ state
\footnote{The GHZ state considered in Ref.~\cite{Lloyd2000} is $\vert \psi\rangle=(1/\sqrt{2})
\left(|\uparrow\rangle_1 |\uparrow\rangle_2 |\uparrow\rangle_3
- |\downarrow\rangle_1 |\downarrow\rangle_2 |\downarrow\rangle_3 \right)$, for which the GHZ correlations
$\langle \psi \vert \sigma_x^1~\sigma_x^2~\sigma_x^3 \vert \psi \rangle = -1$,
$\langle \psi \vert \sigma_x^1~\sigma_y^2~\sigma_y^3 \vert \psi \rangle = +1$,
$\langle \psi \vert \sigma_y^1~\sigma_y^2~\sigma_x^3 \vert \psi \rangle = +1$, and
$\langle \psi \vert \sigma_y^1~\sigma_x^2~\sigma_y^3 \vert \psi \rangle = +1$}, confirming quantum prediction
of GHZ correlations. The perfect GHZ correlations (\ref{xxxc}-\ref{yxyc}) are obtained from
equations (\ref{xxxg}-\ref{yxyg}). In the temporal situation, with the same spirit one can experimentally verify
the three-time temporal correlation functions with respect to the state $|\Phi\rangle$ as follows
\begin{eqnarray}
\label{xxxt}
\langle \Phi \vert \sigma_x(t_3)~\sigma_x(t_2)~\sigma_x(t_1)~|\Phi\rangle &=& -1, \\
\label{xyyt}
\langle \Phi \vert \sigma_x(t_3)~\sigma_y(t_2)~\sigma_y(t_1)~|\Phi\rangle &=& -1, \\
\label{yyxt}
\langle \Phi \vert \sigma_y(t_3)~\sigma_y(t_2)~\sigma_x(t_1)~|\Phi\rangle &=& -1, \\
\label{yxyt}
\langle \Phi \vert \sigma_y(t_3)~\sigma_x(t_2)~\sigma_y(t_1)~|\Phi\rangle &=& +1,
\end{eqnarray}
where the perfect temporal correlations (\ref{xxxt}-\ref{yxyt}) are obtained from the eigenvalue equations (\ref{xxxe}-\ref{yxye}).
NMR technique to measure such temporal correlation functions are already available \cite{Mahesh13,Kavita22}. NMR is one of the most
mature technologies for implementing isolated quantum spin-1/2 dynamics in a magnetic field \cite{NMR1,NMR2}. The NMR
realization of GHZ correlations \cite{Lloyd2000} provides an added motivation for possible experimental verification of our
temporal GHZ result.

\bibliography{TimeGHZ.bib}

\end{document}